\documentclass[twocolumn,prb,showpacs,preprintnumbers,amsmath,amssymb,superscriptaddress]{revtex4}

\usepackage{graphicx}
\usepackage{dcolumn}
\usepackage{bm}

\begin{document}


\title{Charge Sensing of an Artificial $\text{H}^{+}_{2}$ Molecule\\}

\author{M. Pioro-Ladri\`{e}re}
\affiliation{Institute for Microstructural Sciences, National Research Council of Canada, Ottawa, Ontario, Canada K1A~0R6}
\affiliation{D\'epartement de physique and Regroupement qu\'eb\'ecois sur les mat\'eriaux de pointe,
Universit\'e de Sherbrooke, Sherbrooke, Qu\'ebec J1K 2R1, Canada}

\author{M. R. Abolfath}
\author{P. Zawadzki}
\author{J. Lapointe}
\author{S. A. Studenikin}
\author{A. S. Sachrajda}
\author{P. Hawrylak}
\affiliation{Institute for Microstructural Sciences, National Research Council of Canada, Ottawa, Ontario, Canada K1A~0R6}

\date{\today}

\begin{abstract}
We report charge detection studies of a lateral double quantum dot with controllable charge states and tunable tunnel coupling. Using an integrated electrometer, we characterize the equilibrium state of a single electron trapped in the doubled-dot (artificial $\text{H}^{+}_{2}$ molecule) by measuring the average occupation of one dot. We present a model where the electrostatic coupling between the molecule and the sensor is taken into account explicitly. From the measurements, we extract the temperature of the isolated electron and the tunnel coupling energy. It is found that this coupling can be tuned between 0 and 60 $\mu\mathrm{eV}$ in our device.\end{abstract}

\pacs{}

\maketitle

\section{Introduction}
Lateral quantum dots are defined in a two-dimensional electron gas by means of electrical gates. \cite{Kastner93,Ashoori96,Kouwenhoven98,Jacak98} Since they are electrostatically defined, lateral quantum dots are extremely tunable. This advantage is well illustrated by recent achievements involving control over a single electron charge or spin. Experiments have, for example, demonstrated the isolation of single electron in single and double dot systems \cite{ciorga00,Elzerman03,Chan04,Petta04,Huettel05}, the electrical read-out of a single electron spin \cite{Ciorga02,Pioro03,Elzerman04} and the coherent manipulation of the charge state of artificial molecules. \cite{Petta04,Hayashi03} Lateral quantum dots therefore provide the ideal choice for related fundamental studies including those aiming at the implementation of solid state quantum bits. Quantum bits can be naturally encoded either using the spin of a single electron trapped in a quantum dot \cite{Brum97,Loss98} or alternatively, the charge state of a double quantum dot. \cite{Tanamoto00,Blick00,Bayer01,Wiel01} In both approaches, a precise control over the coupling between adjacent dots is important for quantum gate operations.

In this paper, we report charge detection measurements of a double few electron quantum dot containing a tunable number of electrons. By trapping a single electron in the double dot, a two-level system is formed. By analogy to molecular physics, the double-dot can be viewed as an artificial $\text{H}^{+}_{2}$ molecule. \cite{Dybalski05} Following the work of Petta \textit{et al}.\cite{Petta04}, we probe the temperature of this single trapped electron and extract the tunnel coupling energy by charge sensing measurements. We use a quantum point contact (QPC) as a charge sensor and address the effect of the back-action of the sensor on the molecule.

\section{Experimental Techniques}

\begin{figure}
\includegraphics[scale=0.94]{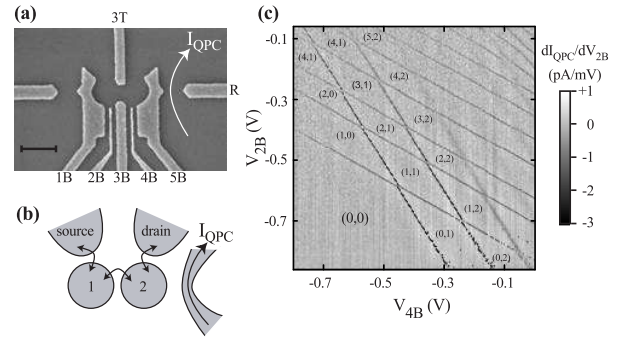}
\caption{(a) SEM picture of the experimental device. Scale bar: 500\,nm; (b) Schematic representation of the different regions shaped by the gates in the 2DEG. Dots 1 and 2 are connected in series to the leads labeled ``source'' and ``drain'' by tunneling barriers. The double-arrows indicate where particle exchange can occur. The QPC current $I_{\mathrm{QPC}}$ flows through a constriction closest to dot 2. Charge sensing is performed by monitoring the transconductance $dI_{\mathrm{QPC}}/dV_{\mathrm{2B}}$; (c) Stability diagram of the double quantum dot obtained by charge sensing. The electron number for each dot is indicated by $\left(N_{1},N_{2}\right)$. The background signal was subtracted for clarity.}
\label{fig1}
\end{figure}

Figure\ \ref{fig1}(a) shows a scanning electron micrograph of our coupled quantum dot device. The device consists of two lateral quantum dots connected in series to leads and coupled capacitively to a QPC. The dots as well as the QPC are defined within the two dimensional electron gas (2DEG) of a GaAs-AlGaAs heterostructure, located 90 nm below the surface (Fig.\ \ref{fig1}(b)). The tunnel barriers indicated by double-arrows in the figure are tuned by negative voltages applied to gates 3T, 1B, 3B and 5B. Gates 2B and 4B may be regarded as ``plungers'' and are used to change the number of electrons. The dot-to-lead tunnel conductance is set to a very small value ($<<2e^{2}/h$) so that the coupled-dot system is well isolated from the leads, whereas the interdot tunnel conductance can be tuned from zero to $2e^{2}/h$. \cite{Livermore96} A negative voltage applied on gate R forms a QPC placed close to the right dot. This QPC, whose conductance is set around $e^{2}/h$, is used as a charge sensor \cite{Field96} for the double-dot system. The device employs a gate layout geometry that permits transport down to the single electron regime. \cite{ciorga00,Elzerman03} Charge detection is performed by monitoring the derivative of the QPC current with respect to the voltage applied on gate 2B. This quantity, $dI_{\mathrm{QPC}}/dV_{\mathrm{2B}}$, the transconductance, is measured using standard lock-in techniques. The device was cooled in a Oxford Kelvinox $400\,\mathrm{\mu W}$ dilution refrigerator with a base temperature of $6\,\mathrm{mK}$.

\section{Results and Discussion}

The occupation numbers $\left(N_{1},N_{2}\right)$, defined as the number of electrons trapped in the left and right dot respectively, are determined by mapping the stability diagram of the double-dot. This can be achieved by measuring the transconductance as a function of the gate voltages $V_{\mathrm{2B}}$ and $V_{\mathrm{4B}}$. Figure \ \ref{fig1}(c) shows a grayscale plot of such a map. For clarity, we subtract the change in the background signal related to the direct capacitive coupling between the gates and the QPC. The dark lines depict a reduced transconductance corresponding to a change in the total electron number in the double-dot. Together these lines form a honey-comb pattern, as expected for a double-dot. \cite{Wiel03} The more horizontal lines correspond to a change of one electron in the left dot. Likewise the more vertical lines occur when the occupation of the right dot changes by one. Away from these lines, each dot is Coulomb blockaded and the system confines a fixed number of electrons. The lack of charging events in the region marked $\left(0,0\right)$ is due to the dot being empty of electrons. \footnote{We checked also that no extra resonances appear as the tunnel barriers to the leads are opened. Data not shown.}

\begin{figure}
\includegraphics[scale=0.9]{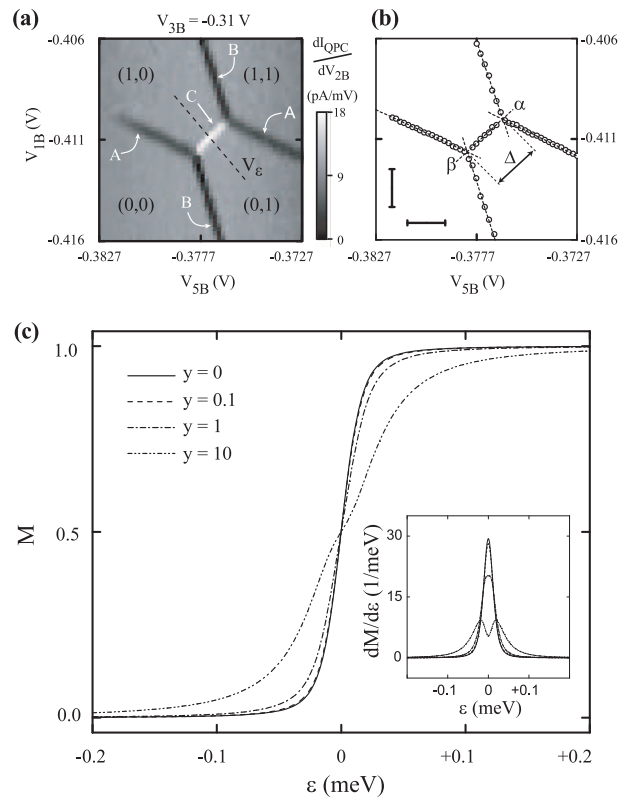}
\caption{(a) Transconductance map near the transition $\left(1,0\right)$ to $\left(0,1\right)$ for $V_{\mathrm{3B}}=-0.31\,\mathrm{V}$; (b) Experimental stability diagram. The positions of the resonances A, B and C from (a) are plotted (open circles). The spacing $\Delta$ between triple points $\alpha$ and $\beta$ is equal to $2t+\gamma$. The scale bars are equal to $300\,\mu\mathrm{eV}$; (c) Average occupation, $M$, of dot 2 as a function of the detuning energy $\epsilon$ calculated using eq.\ \ref{M with backaction} for several values of the back-action parameter $y$. $\Gamma_{r}$, $T_{e}$ and $t$ equal $0.26$, $6$ and $7\,\mu\mathrm{eV}$ for this plot. Inset: Numerical derivative of $M$ with respect to $\epsilon$ of the same curves.}
\label{fig2}
\end{figure}

We now study the equilibrium property of the artificial $\text{H}^{+}_{2}$ molecule. Figure \ \ref{fig2}(a) shows the stability diagram in the relevant regime where the charge states $\left(0,0\right)$, $\left(1,0\right)$, $\left(0,1\right)$ and $\left(1,1\right)$ are accessible. The gate voltages applied to gates 1B and 5B are swept. Three lines of different intensities, marked A, B and C in Fig. \ref{fig2}(a), are present in the greyscale plot. Along resonances A and B, the left and right dot occupation changes from 0 to 1 electron respectively. Because the sensor is more sensitive to changes in the right dot occupation number, the reduced transconductance is more pronounce for resonance B than A. Along resonance C, the electron transfers between dots. Whereas the transconductance dips lower than the background along the resonances A and B, the signal rises to a higher value for line C. This can be easily explained. As we sweep $V_{\mathrm{1B}}$ across resonance A or B in the upward direction, an electron is added to the double dot thereby reducing the QPC conductance. But for resonance C under the same conditions, the electron is transferred from the right to left dots (i.e. away from the sensor), thereby increasing the QPC conductance.

The absence of curvature in the position of peaks A and B implies that the tunnel coupling $t$ between the two dots is very small. Under these conditions the spacing between the points labelled $\alpha$ and $\beta$ in Fig.\ \ref{fig2}(b) is equal to the Coulomb interaction energy $\gamma$ between dots \cite{Ziegler00} which we estimate to be $400\,\mu\mathrm{eV}$. \footnote{The calibration between gate voltage and energy was performed using transport under high bias. \cite{Wiel03} Data not shown.} Since $\gamma$ is large in comparison with thermal energy (which we estimate to $7\,\mu\mathrm{eV}$ from our charge sensing measurements), the configuration in which only one electron is trapped in the double dot system is stable. In this regime, the two charge states $\left(1,0\right)$ and $\left(0,1\right)$ form a two-level system characterized by two orbitals localized in dot 1 and dot 2, with respective energies $\epsilon_{1}$ and $\epsilon_{2}$. Along line C, the energy level of the left dot is aligned with that of the right dot ($\epsilon_{1}=\epsilon_{2}$). Varying $V_{\mathrm{1B}}$ and $V_{\mathrm{5B}}$ along the diagonal perpendicular to line C (labelled ``$V_\epsilon$''in Fig.\ \ref{fig2}(a)) only changes the energy difference $\epsilon=\epsilon_{1}-\epsilon_{2}$ between the two charge states. The lever arm relating gate voltage to detuning can be calibrated experimentally as described below. The tunnel coupling $t$ is tuned in our device via $V_{\mathrm{3B}}$. These two parameters fully characterize the two-level charge system whose Hamiltonian can be written in terms of the Pauli matrices as

\begin{equation}
H = \frac{1}{2}\epsilon\sigma_{z} + \frac{1}{2}2t\sigma_{x}
\label{hamil}
\end{equation}

A quantitative characterization of $\epsilon$ and $t$ is made by measuring the charge sensor signal along the detuning diagonal of the stability diagram at equilibrium. We employ the technique introduced by DiCarlo \textit{et al}. where the average occupation of the right dot, $M$, can be extracted from the QPC current as a function of the detuning between dots. \cite{Dicarlo04} Equivalently, for increased accuracy, we convert the transconductance into the derivative of $M$ with respect to $\epsilon$. Since the effects of the QPC on the two-level system were not addressed in previous works, we fit the data using a refined model, in which the back-action of the QPC on $M$ is taken into account. As we now explain, the model predicts small corrections when the coupling between the double-dot and the sensor is small and a significant change in the line shape of $M$ in the strong coupling regime.

The electrostatic interaction between the trapped electron and the QPC is fundamental to why we can distinguish the two charge states using the sensor signal. This coupling can be characterized by a dephasing rate $\Gamma_{d}$, related to how well the two states can be distinguished. This rate is non zero only if a bias is applied across the QPC and is equal to $\left(\sqrt{I_{1}/e}-\sqrt{I_{2}/e}\right)^{2}$ where $I_{1(2)}$ is the value of the current when the electron is localized in the left (right) dot. \cite{Gurvitz97} The discrete nature of the current through the QPC (resulting in shot-noise) induces fluctuations in the qubit potential causing dephasing. The occupation probability $M$ can be calculated from the equilibrium density matrix of the two-level system. For $\Gamma_{d}=0$, the density matrix is simply the one given by statistical mechanics where the electron trapped in the double-dot is assumed to be in thermal equilibrium with its electromagnetic environment at a temperature $T_{e}$ (which relaxes the molecule to the ground state at a rate $\Gamma_{r}$). Such equilibration can occur through fluctuations in device control parameters such as gate voltage noise. Using the method described in Ref. \cite{Gurvitz03}, the finite temperature equilibrium density matrix in the presence of \textit{both} sensor and reservoir can be calculated analytically. We find the following expression for the average occupation of the right dot
\begin{widetext}
\begin{subequations}
\label{M with backaction}
\begin{eqnarray}
M=\frac{1}{2}-\frac{\Gamma _{xz}\Gamma _{yy}\sin \theta +\left[ \Gamma _{xx}\Gamma _{yy}+\left( \frac{\Omega }{\hbar\Gamma _{r}}\right) ^{2}\right] \cos \theta }{\Gamma _{xz}^{2}\Gamma _{yy}-\left[ \Gamma _{xx}\Gamma _{yy}+\left( \frac{\Omega }{\hbar\Gamma _{r}}\right) ^{2}\right] \Gamma _{zz}}\Gamma _{z}
\\
\Gamma _{xx}=\frac{1}{2}+\frac{y}{2}\cos ^{2}\theta \mathrm{;} \, \Gamma _{yy}=\frac{1}{2}+\frac{y}{2} \mathrm{;} \, \Gamma _{zz}=1+e^{-\beta \Omega }+\frac{y}{2}\sin ^{2}\theta
\\
\Gamma _{xz}=\frac{y}{2}\cos \theta \sin \theta \mathrm{;} \,\Gamma _{z}=\frac{1}{2}-\frac{e^{-\beta \Omega }}{2}
\end{eqnarray}
\end{subequations}
\end{widetext} 
where $\Omega=\sqrt{\epsilon^{2}+4t^{2}}$, $\sin\theta=2t/\Omega$, $\cos\theta=\epsilon/\Omega$, $\beta=1/\mathrm{k}_{\mathrm{B}}T_{e}$, $\mathrm{k}_{\mathrm{B}}$ is the Boltzman constant, $\hbar$ is the reduced Plank constant and $y=\Gamma_{d}/\Gamma_{r}$. 

The ratio $y$ is a quantitative measure of the back-action of the QPC on the molecule. In Fig.\ \ref{fig2}(c), we plot equation \ref{M with backaction} as a function of $\epsilon$ for several values of $y$. For $y=0$, $M$ changes in a step-like manner from 0 to 1 as the detuning is swept across zero. The width of the transition is a combined measure of the tunnel coupling and the thermal energy. For non-zero $\Gamma_{d}$ but small compared to $\Gamma_{r}$, the back-action slightly increases the width. In the regime of strong coupling ($\Gamma_{d}>>\Gamma_{r}$), the back-action strongly distorts the average dot occupation. Note, for example, the formation of a double peak feature in the inset of Fig.\ \ref{fig2}(c) which occurs under conditions of small detuning between the dots and strong coupling to the QPC sensor related to the formation of a plateau of $M=1/2$ in this regime.

\begin{figure}
\includegraphics[scale=1.0]{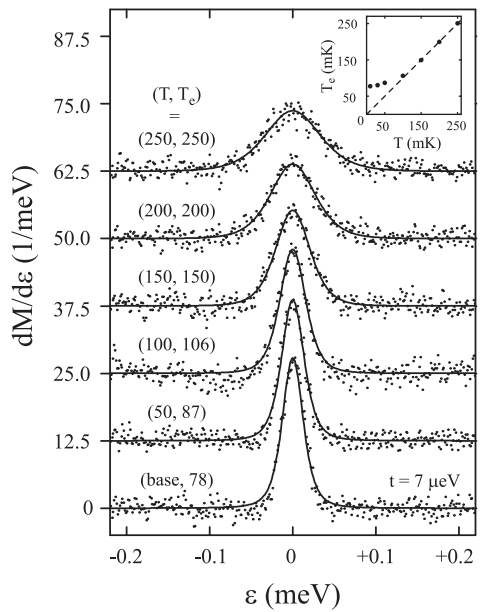}
\caption{Temperature dependence of the $dM/d\epsilon$ peak along the detuning line. $V_{\mathrm{3B}}=-0.31\,\mathrm{V}$. For each trace, the dilution refrigerator and electronic temperature ($T$,$T_{e}$) are indicated in mK. The solid lines are fit of eq. \ref{M with backaction} to the data points assuming $t=7\,\mu\mathrm{eV}$ and $y=0$. Traces are shifted vertically. Inset: $T_{e}$ extracted from the fits plotted as a function of $T$. The line $T_{e}=T$ is also shown (dashed).}
\label{fig3}
\end{figure}

For the experimental conditions used in the present study, we estimate $\Gamma_{d}$ to be approximately 0.12 MHz. \footnote{The voltage bias across the QPC was $100\,\mu\mathrm{eV}$ resulting in a current $I_{1}\approx3.874\,\mathrm{nA}$. The sensitivity of the QPC was such that $I_{1}-I_{2}=17\,\mathrm{pA}$, yielding  $\Gamma_{d}\approx0.12\,\mathrm{MHz}$} Using the relaxation rate measured in Ref. \cite{Petta04} on a similar device ($\Gamma_{r}\approx62.5\,\mathrm{MHz}$), we get $y=0.002$, which is in the weak coupling regime. The modification caused by the back-action to the bare transition (case $y=0$) can be estimated in this regime by expanding $M$ up to first order in y. This modification can be neglected for our estimated value of y since it gives correction of the order of 0.5\%, smaller than our experimental resolution. We therefore fit the data using equation \ref{M with backaction} with $y$ set to 0.

\begin{figure}
\includegraphics[scale=1.0]{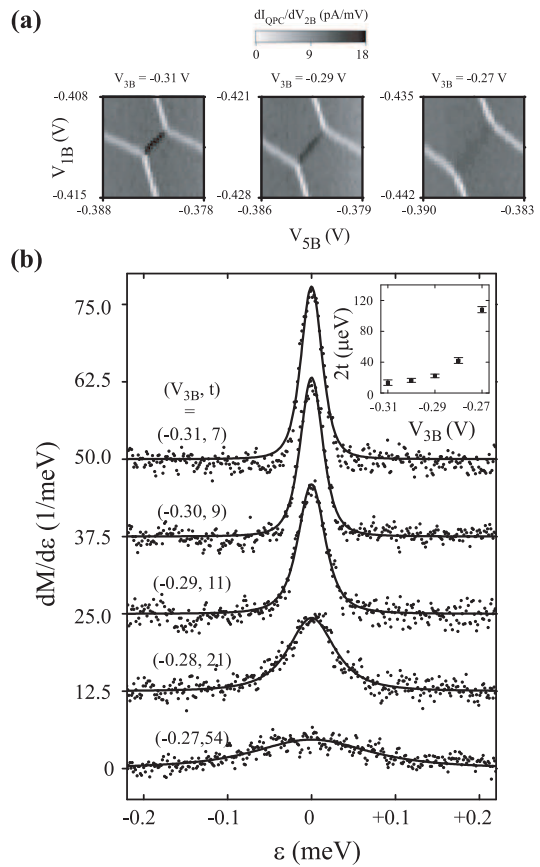}
\caption{(a) Experimental stability diagram obtained for 3 different values of gate voltage $V_{\mathrm{3B}}$; (b) Dependance of the $dM/d\epsilon$ peak with gate voltage. For each trace, the gate voltage and the tunnel coupling ($V_{\mathrm{3B}}$,$t$) are indicated in $\mathrm{V}$ and $\mu\mathrm{eV}$ respectively. The solid lines are fit to the data assuming $T_{e}=78\,\mathrm{mK}$  and $y=0$. Traces are shifted vertically. Inset. $2t$ vs. $V_{\mathrm{3B}}$ extracted from the fits.}
\label{fig4}
\end{figure}

In order to characterize the dependence of the tunnel coupling on gate voltage, we must first estimate the electronic temperature. Figure\ \ref{fig3} shows the effect of increasing the dilution refrigerator temperature $T$ on the width of the resonance. Although for these data the tunnel coupling is weak, its exact value cannot be deduced from the charging diagram. By assuming that for temperatures above 150 mK, the electron is well thermalized to the refrigerator (i.e. $T_{e}=T$), we can calibrate the lever-arm relating the voltage along the detuning diagonal to energy and extract the precise value of $t$. We find that a tunnel coupling of $7\pm1\,\mu\mathrm{eV}$ provides the optimal fit to the data. We then use this calibration to extract the electronic temperature over the complete temperature range. In the inset, we plot $T_{e}$ vs. $T$ obtained from the fits. The electronic temperature at base was found to be 78 mK (nb. without this procedure, if $t$ was simply assumed to be zero the fit would give a value of 101 mK). For comparison the electronic temperature of the 2DEG in the leads was estimated to be approximately 30 mK from activation measurements of the quantum Hall effect whilst the temperature extracted from Coulomb blockade peaks was around 50 mK. This confirms the hypothesis that the more isolated an electron is the more difficult it is to cool. Notice that the signal to noise ratio is high at low temperatures. As a result, even if the back-action parameter $y$ is very small, the changes in $M$ can still be detected.

We now calibrate the gate voltage dependence of the tunnel coupling. Figure \ref{fig4}(a) shows the evolution of the charging diagram as the voltage on gate 3B is made less negative. The tunnel coupling increases leading to an increase in peak spacing and the curvature in the charging diagram. Fig. \ref{fig4}(b) shows the corresponding broadening of the $dM/d\epsilon$ peak. Again, fits to the model assuming an initial tunnel coupling of $7\,\mu\mathrm{eV}$ with a base temperature $T_{e}$ of 78 mK are good. In the inset, we plotted the value of $2t$ extracted for various gate voltages. Over the span in gate voltage used, the tunnel coupling ranged from $7$ to $60\,\mathrm{\mu eV}$. These values are in the same range of those previously reported in the literature. \cite{Petta04,Huettel05,Hayashi03,Wiel03,Dicarlo04} The dependence of the tunnel coupling on the gate voltage $V_{3B}$ is approximately exponential. This suggests that the gate primarily affects the height (and/or the width) of the tunnel barrier of the double-dot potential landscape.

\section{Conclusions}

In conclusion, we have isolated a single electron in a lateral double quantum dot (artificial $\text{H}^{+}_{2}$ molecule) using a QPC as an integrated electrometer with a sensitivity smaller than the elementary charge $e$. The detuning energy and the quantum mechanical amplitude for the electron to tunnel from one dot to the other were characterized for different gate voltages quantitatively from static charge sensing measurements. The electrostatic coupling between the sensor and the double-dot perturbed the state of the electron. A strong dephasing rate associated to the measurement process could significantly modify relevant quantities including the average individual dot occupation probabilities. In our study, the QPC was biased in the weak coupling regime for which we found that this back-action could be ignored. Interestingly, our model suggests that back-action effects could be amplified by increasing the voltage bias across the sensor. \footnote{From the Laudauer formula, the QPC current $I$ is equal to $e^2TV/h$ where $T$ is the transmission probability and $V$ the voltage bias. The electrostatic coupling between the double dot and the sensor affects $T$. In terms of these parameters, $\Gamma_d=\left(\sqrt{T_{1}/e}-\sqrt{T_{2}/e}\right)^{2}e\left|V\right|/h$.} This will be the subject of further studies.

\begin{acknowledgments}
We wish to acknowledge very useful discussions with A. Blais and A. Clerk. M.P.-L. acknowledges assistance from NSERC and FQRNT. A.S.S and P.H. acknowledges support from CIAR.
\end{acknowledgments}


\begin{thebibliography}{27}
\expandafter\ifx\csname natexlab\endcsname\relax\def\natexlab#1{#1}\fi
\expandafter\ifx\csname bibnamefont\endcsname\relax
  \def\bibnamefont#1{#1}\fi
\expandafter\ifx\csname bibfnamefont\endcsname\relax
  \def\bibfnamefont#1{#1}\fi
\expandafter\ifx\csname citenamefont\endcsname\relax
  \def\citenamefont#1{#1}\fi
\expandafter\ifx\csname url\endcsname\relax
  \def\url#1{\texttt{#1}}\fi
\expandafter\ifx\csname urlprefix\endcsname\relax\def\urlprefix{URL }\fi
\providecommand{\bibinfo}[2]{#2}
\providecommand{\eprint}[2][]{\url{#2}}

\bibitem[{\citenamefont{Kastner}(1993)}]{Kastner93}
\bibinfo{author}{\bibfnamefont{M.~A.} \bibnamefont{Kastner}},
  \bibinfo{journal}{Phys. Today} \textbf{\bibinfo{volume}{1}},
  \bibinfo{pages}{24} (\bibinfo{year}{1993}).

\bibitem[{\citenamefont{Ashoori}(1996)}]{Ashoori96}
\bibinfo{author}{\bibfnamefont{R.~C.} \bibnamefont{Ashoori}},
  \bibinfo{journal}{Nature} \textbf{\bibinfo{volume}{379}},
  \bibinfo{pages}{413} (\bibinfo{year}{1996}).

\bibitem[{\citenamefont{Kouwenhoven and Marcus}(June 1998)}]{Kouwenhoven98}
\bibinfo{author}{\bibfnamefont{L.~P.} \bibnamefont{Kouwenhoven}}
  \bibnamefont{and} \bibinfo{author}{\bibfnamefont{C.~M.}
  \bibnamefont{Marcus}}, \bibinfo{journal}{Phys. World}  (\bibinfo{year}{June
  1998}).

\bibitem[{\citenamefont{Jacak et~al.}(1998)\citenamefont{Jacak, Hawrylak, and
  Wojs}}]{Jacak98}
\bibinfo{author}{\bibfnamefont{L.}~\bibnamefont{Jacak}},
  \bibinfo{author}{\bibfnamefont{P.}~\bibnamefont{Hawrylak}}, \bibnamefont{and}
  \bibinfo{author}{\bibfnamefont{A.}~\bibnamefont{Wojs}},
  \emph{\bibinfo{title}{Quantum Dots}} (\bibinfo{publisher}{Springer-Verlag},
  \bibinfo{address}{Berlin}, \bibinfo{year}{1998}).

\bibitem[{\citenamefont{Ciorga et~al.}(2000)\citenamefont{Ciorga, Sachrajda,
  Hawrylak, Gould, Zawadzki, Jullian, Feng, and Wasilewski}}]{ciorga00}
\bibinfo{author}{\bibfnamefont{M.}~\bibnamefont{Ciorga}},
  \bibinfo{author}{\bibfnamefont{A.~S.} \bibnamefont{Sachrajda}},
  \bibinfo{author}{\bibfnamefont{P.}~\bibnamefont{Hawrylak}},
  \bibinfo{author}{\bibfnamefont{C.}~\bibnamefont{Gould}},
  \bibinfo{author}{\bibfnamefont{P.}~\bibnamefont{Zawadzki}},
  \bibinfo{author}{\bibfnamefont{S.}~\bibnamefont{Jullian}},
  \bibinfo{author}{\bibfnamefont{Y.}~\bibnamefont{Feng}}, \bibnamefont{and}
  \bibinfo{author}{\bibfnamefont{Z.}~\bibnamefont{Wasilewski}},
  \bibinfo{journal}{Phys. Rev. B} \textbf{\bibinfo{volume}{61}},
  \bibinfo{pages}{R16315} (\bibinfo{year}{2000}).

\bibitem[{\citenamefont{Elzerman et~al.}(2003)\citenamefont{Elzerman, Hanson,
  Greidanus, Beveren, Franceschi, Vandersypen, Tarucha, and
  Kouwenhoven}}]{Elzerman03}
\bibinfo{author}{\bibfnamefont{J.~M.} \bibnamefont{Elzerman}},
  \bibinfo{author}{\bibfnamefont{R.}~\bibnamefont{Hanson}},
  \bibinfo{author}{\bibfnamefont{J.~S.} \bibnamefont{Greidanus}},
  \bibinfo{author}{\bibfnamefont{L.~H. W.~V.} \bibnamefont{Beveren}},
  \bibinfo{author}{\bibfnamefont{S.~D.} \bibnamefont{Franceschi}},
  \bibinfo{author}{\bibfnamefont{L.~M.~K.} \bibnamefont{Vandersypen}},
  \bibinfo{author}{\bibfnamefont{S.}~\bibnamefont{Tarucha}}, \bibnamefont{and}
  \bibinfo{author}{\bibfnamefont{L.~P.} \bibnamefont{Kouwenhoven}},
  \bibinfo{journal}{Phys. Rev. B} \textbf{\bibinfo{volume}{67}},
  \bibinfo{pages}{161308} (\bibinfo{year}{2003}).

\bibitem[{\citenamefont{Petta et~al.}(2004)\citenamefont{Petta, Johnson,
  Marcus, Hanson, and Gossard}}]{Petta04}
\bibinfo{author}{\bibfnamefont{J.~R.} \bibnamefont{Petta}},
  \bibinfo{author}{\bibfnamefont{A.~C.} \bibnamefont{Johnson}},
  \bibinfo{author}{\bibfnamefont{C.~M.} \bibnamefont{Marcus}},
  \bibinfo{author}{\bibfnamefont{M.~P.} \bibnamefont{Hanson}},
  \bibnamefont{and} \bibinfo{author}{\bibfnamefont{A.~C.}
  \bibnamefont{Gossard}}, \bibinfo{journal}{Phys. Rev. Lett.}
  \textbf{\bibinfo{volume}{93}}, \bibinfo{pages}{186802}
  (\bibinfo{year}{2004}).

\bibitem[{\citenamefont{Huettel et~al.}(2005)\citenamefont{Huettel, Ludwig,
  Eberl, and Kotthaus}}]{Huettel05}
\bibinfo{author}{\bibfnamefont{A.~K.} \bibnamefont{Huettel}},
  \bibinfo{author}{\bibfnamefont{S.}~\bibnamefont{Ludwig}},
  \bibinfo{author}{\bibfnamefont{K.}~\bibnamefont{Eberl}}, \bibnamefont{and}
  \bibinfo{author}{\bibfnamefont{J.~P.} \bibnamefont{Kotthaus}},
  \bibinfo{journal}{cond-mat/0501012}  (\bibinfo{year}{2005}).

\bibitem[{\citenamefont{Chan et~al.}(2004)\citenamefont{Chan, Fallahi, Vidan,
  Westervelt, Hanson, and Gossard}}]{Chan04}
\bibinfo{author}{\bibfnamefont{I.}~\bibnamefont{Chan}},
  \bibinfo{author}{\bibfnamefont{P.}~\bibnamefont{Fallahi}},
  \bibinfo{author}{\bibfnamefont{A.}~\bibnamefont{Vidan}},
  \bibinfo{author}{\bibfnamefont{R.}~\bibnamefont{Westervelt}},
  \bibinfo{author}{\bibfnamefont{M.}~\bibnamefont{Hanson}}, \bibnamefont{and}
  \bibinfo{author}{\bibfnamefont{A.}~\bibnamefont{Gossard}},
  \bibinfo{journal}{Nanotechnology} \textbf{\bibinfo{volume}{15}},
  \bibinfo{pages}{609} (\bibinfo{year}{2004}).

\bibitem[{\citenamefont{Ciorga et~al.}(2002)\citenamefont{Ciorga, Wensauer,
  Pioro-Ladriere, Korkusinski, Kyriakidis, Sachrajda, and Hawrylak}}]{Ciorga02}
\bibinfo{author}{\bibfnamefont{M.}~\bibnamefont{Ciorga}},
  \bibinfo{author}{\bibfnamefont{A.}~\bibnamefont{Wensauer}},
  \bibinfo{author}{\bibfnamefont{M.}~\bibnamefont{Pioro-Ladriere}},
  \bibinfo{author}{\bibfnamefont{M.}~\bibnamefont{Korkusinski}},
  \bibinfo{author}{\bibfnamefont{J.}~\bibnamefont{Kyriakidis}},
  \bibinfo{author}{\bibfnamefont{A.~S.} \bibnamefont{Sachrajda}},
  \bibnamefont{and} \bibinfo{author}{\bibfnamefont{P.}~\bibnamefont{Hawrylak}},
  \bibinfo{journal}{Phys. Rev. Lett.} \textbf{\bibinfo{volume}{88}},
  \bibinfo{pages}{256804} (\bibinfo{year}{2002}).

\bibitem[{\citenamefont{Pioro-Ladri{\`e}re
  et~al.}(2003)\citenamefont{Pioro-Ladri{\`e}re, Ciorga, Lapointe, Zawadzki,
  Korkusiski, Hawrylak, and Sachrajda}}]{Pioro03}
\bibinfo{author}{\bibfnamefont{M.}~\bibnamefont{Pioro-Ladri{\`e}re}},
  \bibinfo{author}{\bibfnamefont{M.}~\bibnamefont{Ciorga}},
  \bibinfo{author}{\bibfnamefont{J.}~\bibnamefont{Lapointe}},
  \bibinfo{author}{\bibfnamefont{P.}~\bibnamefont{Zawadzki}},
  \bibinfo{author}{\bibfnamefont{M.}~\bibnamefont{Korkusiski}},
  \bibinfo{author}{\bibfnamefont{P.}~\bibnamefont{Hawrylak}}, \bibnamefont{and}
  \bibinfo{author}{\bibfnamefont{A.~S.} \bibnamefont{Sachrajda}},
  \bibinfo{journal}{Phys. Rev. Lett.} \textbf{\bibinfo{volume}{91}},
  \bibinfo{pages}{026803} (\bibinfo{year}{2003}).

\bibitem[{\citenamefont{Elzerman et~al.}(2004)\citenamefont{Elzerman, Hanson,
  Beveren, Witkamp, Vandersypen, and Kouwenhoven}}]{Elzerman04}
\bibinfo{author}{\bibfnamefont{J.~M.} \bibnamefont{Elzerman}},
  \bibinfo{author}{\bibfnamefont{R.}~\bibnamefont{Hanson}},
  \bibinfo{author}{\bibfnamefont{L.~H. W.~V.} \bibnamefont{Beveren}},
  \bibinfo{author}{\bibfnamefont{B.}~\bibnamefont{Witkamp}},
  \bibinfo{author}{\bibfnamefont{L.~M.~K.} \bibnamefont{Vandersypen}},
  \bibnamefont{and} \bibinfo{author}{\bibfnamefont{L.~P.}
  \bibnamefont{Kouwenhoven}}, \bibinfo{journal}{Nature}
  \textbf{\bibinfo{volume}{430}}, \bibinfo{pages}{431} (\bibinfo{year}{2004}).

\bibitem[{\citenamefont{Hayashi et~al.}(2003)\citenamefont{Hayashi, Fujisawa,
  Cheong, Jeong, and Hirayama}}]{Hayashi03}
\bibinfo{author}{\bibfnamefont{T.}~\bibnamefont{Hayashi}},
  \bibinfo{author}{\bibfnamefont{T.}~\bibnamefont{Fujisawa}},
  \bibinfo{author}{\bibfnamefont{H.~D.} \bibnamefont{Cheong}},
  \bibinfo{author}{\bibfnamefont{Y.~H.} \bibnamefont{Jeong}}, \bibnamefont{and}
  \bibinfo{author}{\bibfnamefont{Y.}~\bibnamefont{Hirayama}},
  \bibinfo{journal}{Phys. Rev. Lett.} \textbf{\bibinfo{volume}{91}},
  \bibinfo{pages}{226804} (\bibinfo{year}{2003}).

\bibitem[{\citenamefont{Brum and Hawrylak}(1997)}]{Brum97}
\bibinfo{author}{\bibfnamefont{J.~A.} \bibnamefont{Brum}} \bibnamefont{and}
  \bibinfo{author}{\bibfnamefont{P.}~\bibnamefont{Hawrylak}},
  \bibinfo{journal}{Supperlattices Microstruct.} \textbf{\bibinfo{volume}{22}},
  \bibinfo{pages}{431} (\bibinfo{year}{1997}).

\bibitem[{\citenamefont{Loss and DiVincenzo}(1998)}]{Loss98}
\bibinfo{author}{\bibfnamefont{D.}~\bibnamefont{Loss}} \bibnamefont{and}
  \bibinfo{author}{\bibfnamefont{D.~P.} \bibnamefont{DiVincenzo}},
  \bibinfo{journal}{Phys. Rev. A} \textbf{\bibinfo{volume}{57}},
  \bibinfo{pages}{120} (\bibinfo{year}{1998}).

\bibitem[{\citenamefont{Tanamoto}(2000)}]{Tanamoto00}
\bibinfo{author}{\bibfnamefont{T.}~\bibnamefont{Tanamoto}},
  \bibinfo{journal}{Phys. Rev. A} \textbf{\bibinfo{volume}{61}},
  \bibinfo{pages}{022305} (\bibinfo{year}{2000}).

\bibitem[{\citenamefont{Blick and Lorenz}(2000)}]{Blick00}
\bibinfo{author}{\bibfnamefont{R.}~\bibnamefont{Blick}} \bibnamefont{and}
  \bibinfo{author}{\bibfnamefont{H.}~\bibnamefont{Lorenz}},
  \bibinfo{journal}{Proc. IEEE Int. Symp. Circuits and Systems} pp.
  \bibinfo{pages}{II--245} (\bibinfo{year}{2000}).

\bibitem[{\citenamefont{Bayer et~al.}(2001)\citenamefont{Bayer, Hawrylak,
  Hinzer, Fafard, Korkusinski, Wasilewski, Stern, and Forchel}}]{Bayer01}
\bibinfo{author}{\bibfnamefont{M.}~\bibnamefont{Bayer}},
  \bibinfo{author}{\bibfnamefont{P.}~\bibnamefont{Hawrylak}},
  \bibinfo{author}{\bibfnamefont{K.}~\bibnamefont{Hinzer}},
  \bibinfo{author}{\bibfnamefont{S.}~\bibnamefont{Fafard}},
  \bibinfo{author}{\bibfnamefont{M.}~\bibnamefont{Korkusinski}},
  \bibinfo{author}{\bibfnamefont{Z.~R.} \bibnamefont{Wasilewski}},
  \bibinfo{author}{\bibfnamefont{O.}~\bibnamefont{Stern}}, \bibnamefont{and}
  \bibinfo{author}{\bibfnamefont{A.}~\bibnamefont{Forchel}},
  \bibinfo{journal}{Science} \textbf{\bibinfo{volume}{291}},
  \bibinfo{pages}{451} (\bibinfo{year}{2001}).

\bibitem[{\citenamefont{der Wiel et~al.}(2001)\citenamefont{der Wiel, Fujisawa,
  Tarucha, and Kouwenhoven}}]{Wiel01}
\bibinfo{author}{\bibfnamefont{W.~V.} \bibnamefont{der Wiel}},
  \bibinfo{author}{\bibfnamefont{T.}~\bibnamefont{Fujisawa}},
  \bibinfo{author}{\bibfnamefont{S.}~\bibnamefont{Tarucha}}, \bibnamefont{and}
  \bibinfo{author}{\bibfnamefont{L.}~\bibnamefont{Kouwenhoven}},
  \bibinfo{journal}{Jpn J. App. Phys.} \textbf{\bibinfo{volume}{40}},
  \bibinfo{pages}{2100} (\bibinfo{year}{2001}).

\bibitem[{\citenamefont{Dybalski and Hawrylak}(2005)}]{Dybalski05}
\bibinfo{author}{\bibfnamefont{W.}~\bibnamefont{Dybalski}} \bibnamefont{and}
  \bibinfo{author}{\bibfnamefont{P.}~\bibnamefont{Hawrylak}},
  \bibinfo{journal}{cond-mat/0502161}  (\bibinfo{year}{2005}).

\bibitem[{\citenamefont{Livermore et~al.}(1996)\citenamefont{Livermore, Crouch,
  Westervelt, Campman, and Gossard}}]{Livermore96}
\bibinfo{author}{\bibfnamefont{C.}~\bibnamefont{Livermore}},
  \bibinfo{author}{\bibfnamefont{C.~H.} \bibnamefont{Crouch}},
  \bibinfo{author}{\bibfnamefont{R.~M.} \bibnamefont{Westervelt}},
  \bibinfo{author}{\bibfnamefont{K.~L.} \bibnamefont{Campman}},
  \bibnamefont{and} \bibinfo{author}{\bibfnamefont{A.~C.}
  \bibnamefont{Gossard}}, \bibinfo{journal}{Science}
  \textbf{\bibinfo{volume}{274}}, \bibinfo{pages}{1332} (\bibinfo{year}{1996}).

\bibitem[{\citenamefont{Field et~al.}(1996)\citenamefont{Field, Smith, Pepper,
  Ritchie, Frost, Jones, and Hasko}}]{Field96}
\bibinfo{author}{\bibfnamefont{M.}~\bibnamefont{Field}},
  \bibinfo{author}{\bibfnamefont{C.~G.} \bibnamefont{Smith}},
  \bibinfo{author}{\bibfnamefont{M.}~\bibnamefont{Pepper}},
  \bibinfo{author}{\bibfnamefont{D.~A.} \bibnamefont{Ritchie}},
  \bibinfo{author}{\bibfnamefont{J.~E.~F.} \bibnamefont{Frost}},
  \bibinfo{author}{\bibfnamefont{G.~A.~C.} \bibnamefont{Jones}},
  \bibnamefont{and} \bibinfo{author}{\bibfnamefont{D.~G.} \bibnamefont{Hasko}},
  \bibinfo{journal}{Phys. Rev. Lett.} \textbf{\bibinfo{volume}{70}},
  \bibinfo{pages}{1311} (\bibinfo{year}{1996}).

\bibitem[{\citenamefont{der Wiel et~al.}(2003)\citenamefont{der Wiel,
  Franceschi, Elzerman, Fujisawa, Tarucha, and Kouwenhoven}}]{Wiel03}
\bibinfo{author}{\bibfnamefont{W.~G.~V.} \bibnamefont{der Wiel}},
  \bibinfo{author}{\bibfnamefont{S.~D.} \bibnamefont{Franceschi}},
  \bibinfo{author}{\bibfnamefont{J.~M.} \bibnamefont{Elzerman}},
  \bibinfo{author}{\bibfnamefont{T.}~\bibnamefont{Fujisawa}},
  \bibinfo{author}{\bibfnamefont{S.}~\bibnamefont{Tarucha}}, \bibnamefont{and}
  \bibinfo{author}{\bibfnamefont{L.~P.} \bibnamefont{Kouwenhoven}},
  \bibinfo{journal}{Rev. Mod. Phys.} \textbf{\bibinfo{volume}{75}},
  \bibinfo{pages}{1} (\bibinfo{year}{2003}).

\bibitem[{\citenamefont{Ziegler et~al.}(2000)\citenamefont{Ziegler, Bruder, and
  Schoeller}}]{Ziegler00}
\bibinfo{author}{\bibfnamefont{R.}~\bibnamefont{Ziegler}},
  \bibinfo{author}{\bibfnamefont{C.}~\bibnamefont{Bruder}}, \bibnamefont{and}
  \bibinfo{author}{\bibfnamefont{H.}~\bibnamefont{Schoeller}},
  \bibinfo{journal}{Phys. Rev. B} \textbf{\bibinfo{volume}{62}},
  \bibinfo{pages}{1961} (\bibinfo{year}{2000}).

\bibitem[{\citenamefont{DiCarlo et~al.}(2004)\citenamefont{DiCarlo, Lynch,
  Johnson, Childress, Crockett, Marcus, Hanson, and Gossard}}]{Dicarlo04}
\bibinfo{author}{\bibfnamefont{L.}~\bibnamefont{DiCarlo}},
  \bibinfo{author}{\bibfnamefont{H.~J.} \bibnamefont{Lynch}},
  \bibinfo{author}{\bibfnamefont{A.~C.} \bibnamefont{Johnson}},
  \bibinfo{author}{\bibfnamefont{L.~I.} \bibnamefont{Childress}},
  \bibinfo{author}{\bibfnamefont{K.}~\bibnamefont{Crockett}},
  \bibinfo{author}{\bibfnamefont{C.~M.} \bibnamefont{Marcus}},
  \bibinfo{author}{\bibfnamefont{M.~P.} \bibnamefont{Hanson}},
  \bibnamefont{and} \bibinfo{author}{\bibfnamefont{A.~C.}
  \bibnamefont{Gossard}}, \bibinfo{journal}{Phys. Rev. Lett.}
  \textbf{\bibinfo{volume}{92}}, \bibinfo{pages}{226801}
  (\bibinfo{year}{2004}).

\bibitem[{\citenamefont{Gurvitz}(1997)}]{Gurvitz97}
\bibinfo{author}{\bibfnamefont{S.~A.} \bibnamefont{Gurvitz}},
  \bibinfo{journal}{Phys. Rev. B} \textbf{\bibinfo{volume}{56}},
  \bibinfo{pages}{15215} (\bibinfo{year}{1997}).

\bibitem[{\citenamefont{Gurvitz et~al.}(2003)\citenamefont{Gurvitz, Fedichkin,
  Mozyrsky, and Berman}}]{Gurvitz03}
\bibinfo{author}{\bibfnamefont{S.~A.} \bibnamefont{Gurvitz}},
  \bibinfo{author}{\bibfnamefont{L.}~\bibnamefont{Fedichkin}},
  \bibinfo{author}{\bibfnamefont{D.}~\bibnamefont{Mozyrsky}}, \bibnamefont{and}
  \bibinfo{author}{\bibfnamefont{G.}~\bibnamefont{Berman}},
  \bibinfo{journal}{Phys. Rev. Lett.} \textbf{\bibinfo{volume}{91}},
  \bibinfo{pages}{066801} (\bibinfo{year}{2003}).

\end{thebibliography}

\end{document}